# A WARMER AND WETTER SOLUTION FOR EARLY MARS AND THE CHALLENGES WITH TRANSIENT WARMING

Ramses M. Ramirez[i],[ii]

[i]Carl Sagan Institute, Cornell University, Ithaca, NY, USA 14853

[ii]Cornell Center of Astrophysics and Planetary Science, Cornell University, Ithaca, NY, USA 14850

Corresponding author (Ramses Ramirez) email: rmr277@cornell.edu

Cornell University

304 Space Sciences Building

Ithaca, NY, USA 14853

ph: 1(480)-296-6477

Keywords: Mars, atmosphere; Mars, climate; Atmospheres, evolution

# ABSTRACT

The climate of early Mars has been hotly debated for decades. Although most investigators believe that the geology indicates the presence of surface water, disagreement has persisted regarding how warm and wet the surface must have been and how long such conditions may have existed. Although the geologic evidence is most easily explained by a persistently warm climate, the perceived difficulty that climate models have in generating warm surface conditions has seeded various models that assume a cold and glaciated early Mars punctuated by transient warming episodes. However, I use a single-column radiative-convective climate model to show that it is relatively more straightforward to satisfy warm and relatively non-glaciated early Mars conditions, requiring only ~1% $H_2$ and 3 bar $CO_2$ or ~20% $H_2$ and 0.55 bar $CO_2$. In contrast, the reflectivity of surface ice greatly increases the difficulty to transiently warm an initially frozen surface. Surface pressure thresholds required for warm conditions increase ~10 – 60% for transient warming models, depending on ice cover fraction. No warm solution is possible for ice cover fractions exceeding 40%, 70%, and 85% for mixed snow/ice and 25%, 35%, and 49% for fresher snow/ice at $H_2$ concentrations of 3%, 10%, and 20%, respectively. If high temperatures (298 – 323 K) were required to produce the observed surface clay amounts on a transiently warm early Mars (Bishop et al), I show that such temperatures would have required surface pressures that exceed available paleopressure constraints for nearly all $H_2$ concentrations considered (1 – 20%). I then argue that a warm and semi-arid climate remains the simplest and most logical solution to Mars paleoclimate.

## 1. INTRODUCTION

The climate of early Mars remains an intriguing mystery. The surface geology, including the presence of immense valley networks at the Noachian-Hesperian transition (hereafter, older valleys), suggests that stable liquid water flowed on its surface ~3.8 – 3.6 Ga (e.g. Pollack et al., 1987). Disagreement persists regarding how warm and wet the surface must have been and how long such conditions would have lasted. In light of the geologic evidence (e.g. Craddock and Howard, 2002), however, arguments for a permanently cold Mars have mostly subsided. Current arguments acknowledge that Mars must have been at least episodically warm (e.g. Kite et al. 2014).

It is critical to note that such proposed episodic warm periods for early Mars should not be equated with the snowball episodes thought to have repeatedly occurred throughout Earth's past, though (Hoffman et al. 1998). In spite of the faint young Sun paradox (e.g. Ramirez, 2016), Earth's climate has been predominantly warm albeit periodically interrupted by transient "cooling" episodes (e.g. Hoffman et al., 1998). In contrast, the episodic warming mechanisms proposed for early Mars argue the opposite, mainly that the planet was predominantly cold, but was interrupted with short ($< \sim$10 Myr) warm periods (e.g. Kite et al. 2014)." Also, unlike proposed Mars transient warming mechanisms, which are all abiotic, life is often proposed as the trigger for Earth snowball events (e.g. Pavlov et al., 2003; Kopp et al., 2005).

Nevertheless, climate scientists have had much difficulty trying to simulate warm conditions be they for a transiently or persistently warm early Mars (e.g. Kasting et al., 1991; Forget et al. 2013; Wordsworth et al. 2013). Climate models have repeatedly shown that greenhouse warming solely provided by $CO_2$ and $H_2O$ vapor is insufficient to generate warm surface conditions (e.g. Kasting, 1991; Wordsworth et al., 2013; Ramirez et al. 2014a; Wordsworth et al. 2017). Supplementing this warming with that from $CO_2$ clouds (Forget and Pierrehumbert 1997) cannot resolve the paradox either (e.g., Kitzmann et al., 2013; 2016). Urata and Toon (2013) suggested that $H_2O$ vapor cirrus clouds composed of 10-micron (or larger) sized particles could have provided the requisite warming. However, these authors assumed cloud particle residence times that are likely too large by a factor of ~20, suggesting that cirrus cloud warming was greatly overestimated (Ramirez and Kasting, 2017). Moreover, sufficient warming would require cirrus cloud coverage to exceed ~75%, even assuming the most ideal conditions, which may be unrealistic for planets with both rising and subsiding air (ibid).

The difficulties in producing warming climates with just $CO_2$ and $H_2O$ vapor have prompted numerous investigators to include the warming of secondary greenhouse gases in $CO_2$-$H_2O$ atmospheres. Although $SO_2$ is a moderately strong greenhouse gas (Postawko and Kuhn, 1986), it 1) forms highly reflective sulfate and sulfur aerosols which reduce greenhouse warming, and 2) it is short-lived as it rains out of the atmosphere once warm conditions are achieved (Tian et al. 2010). Other possibilities include $CH_4$ and $H_2$, both which may have been outgassed from an initially reduced mantle (Grott et al., 2011). In the absence of $CO_2$-$CH_4$ collisions, Ramirez et al. (2014a) showed that upper atmospheric absorption from $CH_4$ would offset that in the troposphere. Once $CO_2$-$CH_4$ collision-induced absorption was included, however, $CH_4$ was found to produce some warming in these atmospheres (Wordsworth et al., 2017). Moreover, Ramirez et al. (2014a) suggested that a reduced early martian mantle on a warm early Mars would have outgassed $H_2$, which absorbs really well in collisions with a background gas, like $N_2$ (Wordsworth and Pierrehumbert, 2013) or $CO_2$. The simulations in Ramirez et al. (2014a) showed that a warm early climate could have been achieved with ~ 3 – 4 bar $CO_2$ and 5% $H_2$ or 1.3 bar $CO_2$ and 20% $H_2$. The latter satisfies one set of $CO_2$ paleopressure constraints (Kite et al. 2014) and another if large amounts of carbonates formed in open lakes (Hu et al. 2015). Wordsworth et al. (2017) had also updated the efficacy of $CO_2$-$H_2$ absorption, finding it even more effective than that assumed in Ramirez et al. (2014a).

However, the warming effect of $CH_4$ is still insufficient to produce warm conditions unless $CH_4$ concentrations exceeded ~10% in a 2-bar $CO_2$ atmosphere (Wordsworth et al. 2017), which would generate anti-greenhouse hazes that cool the climate (Haqq-Misra et al. 2008). This $CO_2$ partial pressure also exceeds available paleopressure atmospheric constraints (Kite et al. 2014; Hu et al. 2015), possibly by more than a factor of 2 (Hu et al. 2015). More importantly, Wordsworth et al. (2017) advocate an "icy highlands" scenario that posits that Mars was predominantly icy and cold, with the main accumulations of water stored as glaciers at high elevations (Wordsworth et al., 2013; 2016). In this cold Mars model, transient warming episodes are possible through external events (e.g. volcanism, impacts, or $CH_4$ bursts) that generate glacial melt that flow down to lower elevations and carve the older valley networks. This process continues until temperatures again fall below the freezing point of water and the build-up of highlands sheets can recommence. Only partial melting, as opposed to complete melting, of the accumulated ice during each episode is required but the planetary albedo of a heavily glaciated cold Mars should be higher than that for a warm Mars model and require higher greenhouse concentrations to achieve warm conditions . However, Wordsworth et al. (2017) had assumed a low surface albedo that would have only been consistent with ice-free (or nearly ice-free) surface conditions and so their computed greenhouse gas partial pressures should be considered too low for their icy highlands scenario.

Here, I update the analysis in Ramirez et al. (2014a) with the updated $CO_2$-$H_2$ and $CO_2$-$CH_4$ collision-induced absorption parameterizations of Wordsworth et al. (2017) and evaluate the greenhouse gas partial pressure requirements for initially warm and cold early Mars states. As in previous studies (e.g. Kasting, 1991; Ramirez et al., 2014a), I assume that a warm early Mars has a surface albedo of ~0.216, nearly identical to that Viking had measured as an average of the values in the near-equatorial regions (Kieffer et al. 1977). I simulate the effect of ice cover by increasing this surface albedo over a range of reasonable values (0.3 – 0.65) and compute the resultant surface temperatures and required greenhouse gas amounts. I explicitly show that a higher surface albedo significantly increases the difficultly to deglaciate a cold early Mars and that resultant surface pressures would exceed known constraints in many cases. I then show that very high surface pressures, exceeding current constraints (Kite et al., 2014; Hu et al., 2015), would also be needed if high temperatures are needed to form observed abundances of surface clays (Bishop et al. 2017). I also update earlier methane calculations (Ramirez et al. 2014a) and assess the potential role of cirrus cloud warming. On the basis of these arguments, I conclude that a warm (possibly semi-arid) early Mars remains the simplest and most logical solution to Martian paleoclimate.

## 2. METHODS

*2.1 Climate model description*

I used a single-column radiative-convective climate model first developed by Kasting et al. (1993) and updated in recent studies (Ramirez et al., 2014a; Ramirez and Kaltenegger, 2016; 2017). The atmosphere is subdivided into 100 vertical logarithmically-spaced layers that extend from the ground to a low pressure at the top of the atmosphere ($5\times10^{-5}$ bar for Mars). Absorbed and emitted radiative fluxes in the stratosphere are assumed to be balanced. Should tropospheric radiative lapse rates exceed their moist adiabatic values, the model relaxes to a moist $H_2O$ adiabat at high temperatures, or to a moist $CO_2$ adiabat when it is cold enough tor $CO_2$ to condense (Kasting, 1991). This defines a convective troposphere which is assumed to be fully-saturated. This last assumption overestimates the actual greenhouse effect, but sensitivity studies at lower relative humidity (50%) both in previous studies (Ramirez et al. 2014a), and in the current study, indicate that errors in mean surface temperatures are small enough for the purposes of this analysis (< ~3 K) given the uncertainties in atmospheric relative humidity on early Mars.

I employ a standard two-stream approximation for both the solar and infrared portions of the radiative code (Toon et al. 1989). This scheme is appropriate for the atmospheres considered here but not for those involving $CO_2$ clouds (Kitzmann et al. 2013). My correlated-*k* coefficients parameterize gaseous absorption across 38 solar spectral intervals spanning from 0.2 to 4.5 microns (~2000 – 50,0000 $cm^{-1}$) and 55 thermal infrared intervals ranging from 0 – 15,000 $cm^{-1}$ (> ~0.66 microns). As in Ramirez and Kasting (2017) and Ramirez et al. (2014a), I used separate 8-term HITRAN and HITEMP coefficients for $CO_2$ and $H_2O$, respectively, truncated at 500 $cm^{-1}$ and 25 $cm^{-1}$, respectively, computed over 8 temperatures (100, 150, 200, 250, 300, 350, 400, 600 K), and 8 pressures ($10^5$ – 100 bar) (Kopparapu et al. 2013; Ramirez et al. 2014ab). For $CH_4$, 4-term thermal infrared HITRAN k-coefficients are derived over 5 temperatures (100, 200, 300, 400, and 600 K) over the same pressures. The near-IR $CH_4$ coefficients for wavelengths less than 1 micron are those derived from Karkoschka (1994). Overlap between gases was computed by convolving the k-coefficients for all gases within each broadband spectral interval.

Following Ramirez et al. (2014a), I parameterize far-wing absorption in the 15-micron band of $CO_2$ utilizing the 4.3-micron region as a proxy (Perrin and Hartman, 1989). In analogous fashion, the BPS water continuum of Paynter and Ramaswamy (2011) is overlain over its region of validity (0 - ~18,0000 $cm^{-1}$). $CO_2$-$CO_2$ collision-induced absorption (CIA) between $CO_2$ molecules is parameterized using the most standard formulation (Gruzka and Boysow, 1997; 1998; Baranov et al. 2004). I parameterize $CO_2$-$H_2$ and $CO_2$-$CH_4$ CIA using the ab initio calculations and data from Wordsworth et al. (2017). I have implemented the $N_2$-$H_2$ CIA scheme from Borysow and Frommhold (1986). I have also included $CH_4$-$CH_4$, $N_2$-$CH_4$, and $N_2$-$N_2$ CIA (Richard et al., 2012) although sensitivity studies (not shown) reveal that their contributions are negligible in these predominantly $CO_2$-$H_2O$-$H_2$ atmospheres. I use standard $H_2O$ vapor Rayleigh scattering coefficients (Vardavas and Carver, 1984) utilizing available data (Bucholtz, 1995; Edlén, 1966). Rayleigh scattering data from $CO_2$ and $CH_4$ come from Allen (1976) and Sneep & Ubachs (2005), respectively, whereas that by $H_2$ comes from Dalgarno and Williams (1962). My model includes the decrease of gravity with altitude, following Ramirez et al. (2014a), which slightly decreases thermal infrared emission. Calculated surface temperatures are very comparable to those computed in Wordsworth et al (2017) Fig. 2 even though the latter assumed subsaturated (80%) tropospheric relative humidity and employed different model assumptions.

*2.2 Climate modeling procedures*

I assumed that the Sun remains fixed at a solar zenith angle of 60º as in previous works (Kasting, 1991; Ramirez et al., 2014a). An average martian solar flux value of 439 W/m$^2$ was assumed for early mars conditions ~3.8 Ga. Unless otherwise specified, most modeled atmospheres are cloud-free and are predominantly composed of $CO_2$ (80 – 95%) with varying amounts of (1-20%) $H_2$ (or 1% $CH_4$ in the sensitivity study of Section 3.5.3). Following Ramirez et al. (2014a), this range of hydrogen concentrations was determined by assuming that 1) volcanic outgassing rates per unit area on early Mars were similar to those of modern Earth (Móntesi and Zuber, 2003) and 2) that the mantle could have been several log units more reduced than Earth's (Grott et al. 2011). Spherical geometry and planetary magnetic fields could have also favored hydrogen retention (Stone and Proga, 2009; Ramirez et al. 2014a). Hydrogen is assumed to escape at the diffusion limit, the fastest escape rate at such concentrations (e.g. Walker, 1977). The remaining background gas is $N_2$.

As in Ramirez et al. (2014a) and Kasting (1991), I adjust the surface albedo of my model to a value (0.216) that yields the mean surface temperature for present day Mars (~218 K) and use this albedo for the baseline calculations that assume an initially warm, virtually ice-free, southern highlands. For an initially glaciated planet consistent with an initially cold Mars, I vary surface albedo over a range of reasonable values (0.3, 0.40, 0.45, 0.5, 0.55, 0.6 0.65). At such albedo values, the surfaces of the glaciated planets in these simulations are assumed to be only partially ice-covered, not globally ice-covered, which is consistent with GCM simulations (e.g. Wordsworth et al. 2013).

Unless otherwise specified (see next section), I employ the method of "forward calculations" (e.g. Kasting, 1991) and use the martian solar flux at 3.8 Ga to compute the resultant tropospheric and stratospheric temperature profiles at each time step. My calculations begin at a low surface pressure (6x10$^{-3}$ bar) and are gradually incremented, with a new mean surface temperature computed at each pressure increment until either above-freezing mean surface temperatures are achieved or the combined effects of $CO_2$ condensation and Rayleigh scattering prevent further warming.

*Cirrus cloud sensitivity study*

I also assess the effect that cirrus cloud warming could have in these $CO_2$-$H_2$ atmospheres, following a suggestion of their potential effect on early Mars climate (Urata and Toon, 2013). For these calculations, I use a similar procedure as that outlined in Ramirez and Kasting (2017) or Rondanelli and Lindzen (2010). I have maximized cloud greenhouse warming by assuming thin cirrus cloud decks (~ 1km) in fully-saturated atmospheres. The number of atmospheric layers is increased to 200 to better resolve these thin clouds. Following previous studies (e.g. Urata and Toon, 2013; Ramirez et al. 2014b; Ramirez and Kasting, 2017), I use Mie optical properties appropriate for 100-micron cloud particles. Wavenumber-dependent optical depths ($\tau$) were computed using a standard equation (equation 1 in Ramirez and Kasting, 2017).

I follow a similar technique as that in previous studies (e.g. Kasting, 1988; Ramirez et al., 2014b; Ramirez and Kasting, 2017) to gauge the radiative effects of cloud forcing. The clouds were then placed at various heights within the convective troposphere. I then used the "inverse calculation" technique which specifies both a constant temperature stratospheric profile and a mean surface temperature to calculate the cloud forcing required to achieve radiative-convective equilibrium at that surface temperature (e.g. Kasting, 1988, 1991). A moist adiabat is integrated upwards until it intersects with the stratospheric temperature profile. In these calculations, I adopt a stratospheric temperature of 155 K and assumed a mean surface temperature of 273 K, corresponding to a warm climate. In contrast to Ramirez and Kasting (2017), I redefine cloud forcing as $F_S - F_{IR}$, with positive values indicating that clouds warm the climate ($F_S$ = the net incoming absorbed solar flux; $F_{IR}$ = the net outgoing infrared flux).

Fractional cloud cover is modeled by averaging cloudy and cloud-free radiative fluxes (e.g. Arking et al., 1996; Ramirez and Kasting, 2017). For example, the fluxes at 50% cloud cover were determined by averaging the respective clear-sky and 100% cloud cover $F_{IR}$ and $F_S$ values. In analogous fashion, $F_{IR}$ and $F_S$ at 25% cloud cover is computed by averaging the corresponding fluxes at 50% and 0% cloud cover. Those at 38% cloud cover were calculated by averaging the corresponding fluxes at 50% and 25%.

# 3. RESULTS

## 3.1 Comparison with previous work and paleopressure constraints

I repeated the analysis in Figure 2 of Ramirez et al. (2014a) using the new $CO_2$-$H_2$ collision-induced absorption data from Wordsworth et al. (2017) (Figure 1). The mean surface temperature of the baseline hydrogen-free (95% $CO_2$, 5% $N_2$) atmosphere was ~ 230 K, which agrees well with previous works (e.g. Tian et al. 2010; Ramirez et al. 2014a; Wordsworth et al., 2017). Whereas Ramirez et al. (2014a) required ~5% $H_2$ (and 3 - 4 bar surface pressure) to achieve above-freezing mean surface temperatures, only ~1% $H_2$ (and 3 bar) is needed to reach 273 K in this study. Wordsworth et al. (2017) had not found this result because the maximum pressure they considered was 2 bar.

Higher $H_2$ concentrations allow the freezing point to be reached at still lower surface pressures. For 3% $H_2$ this threshold was reached at ~ 1.8 bar; for 5% $H_2$ it was reached at ~1.4 bar; at 10% $H_2$, the corresponding pressure was ~1 bar. These calculated pressures and concentrations agree well with those of Wordsworth et al. (2017) in spite of model differences (see Methods). I also performed a calculation at 20% $H_2$ and found that only ~0.7 bar (~0.55 bar $CO_2$) is needed to achieve warm conditions (Figure 1a). At such low $CO_2$ pressures, the missing carbonate problem for a warm early Mars climate (e.g. Wray et al. 2016) becomes less of one.

Thus, whereas Ramirez et al. (2014a) found that concentrations exceeding ~20% $H_2$ were necessary to satisfy the 1.6 bar atmospheric paleopressure limit calculated by Kite et al. (2014), an $fH_2$ above only ~3.5% would satisfy this condition. Those calculations assume a weak soil target strength consistent with river alluvium, however, and target materials akin with bedrock would require still higher pressures (> 5 bar). (ibid). A second paleopressure estimate suggests that the early atmospheric pressure may have been under ~1 bar (Hu et al. 2015), increasing to 1.8 bar if large amounts of carbonation formation had occurred in open lakes (ibid). Indeed, large amounts of carbonate formation would be expected on a warm early Mars with a dense $CO_2$ atmosphere. Nevertheless, hydrogen concentrations exceeding ~10% and ~3%, respectively, satisfy both of these suggested constraints. The effectiveness of warming by $CO_2$-$H_2$ arises from strong absorption in thermal infrared regions where $H_2O$ and $CO_2$ absorb relatively poorly (i.e. Ramirez et al. 2014a).

Planetary albedo generally increases as the hydrogen concentration decreases (Figure 1b). This is because $CO_2$ rayleigh scattering is stronger than that of the remaining gases, and as $H_2$ concentrations increase, overall atmospheric scattering decreases (e.g. Ramirez et al. (2014a). This is also why the $CO_2$-free atmosphere (80% $N_2$, 20% $H_2$) has the lowest planetary albedo at higher pressures (Figure 1b).

*3.2 Limit cycle stability*

I also compare the climate stability of my warm solutions within the context of "limit cycles" (Batalha et al., 2016). This idea suggests that early Mars was predominantly cold and glaciated, punctuated by short intervals (~ 5 – 10 Myr each) of deglaciation and transient warming. Assuming $CO_2$ is transferred efficiently between the surface and the mantle, these limit cycles occur when $CO_2$ outgassing is outpaced by its removal via other processes (e.g. rainfall, weathering, carbonate formation). Ideally, if the weathering rate is linearly proportional to the dissolved [H+] in groundwater, then the dependence of the weathering rate on the $CO_2$ partial pressure ($\beta$) is 0.5 (Menou, 2015; Batalha et al. 2016). With this $\beta$ value, limit cycles occur for $H_2$ concentrations < ~11% (Figure 2). However, terrestrial soil studies measure a weaker fractional order dependence between H+ and weathering rates (e.g., Asolekar et al., 1991), suggesting that $\beta$ is between ~ 0.3 – 0.4 for the range of silicate rocks (e.g. Lasaga, 1984; Schwartzman and Volk, 1989). Moreover, $\beta$ values as low as 0.25 are possible for acidic environments with pH < 5 (Berner, 1992). Such low-pH environments are predicted for the dense $CO_2$-rich atmospheres thought to have characterized early Mars (e.g. Fairen et al. 2010), suggesting that low $\beta$ values may be more appropriate. At a $\beta$ of 0.3 (0.4), only 1% (3%) $H_2$ is needed for warm stable solutions (Figure 2), suggesting that limit cycles may not have occurred on a warm early Mars.

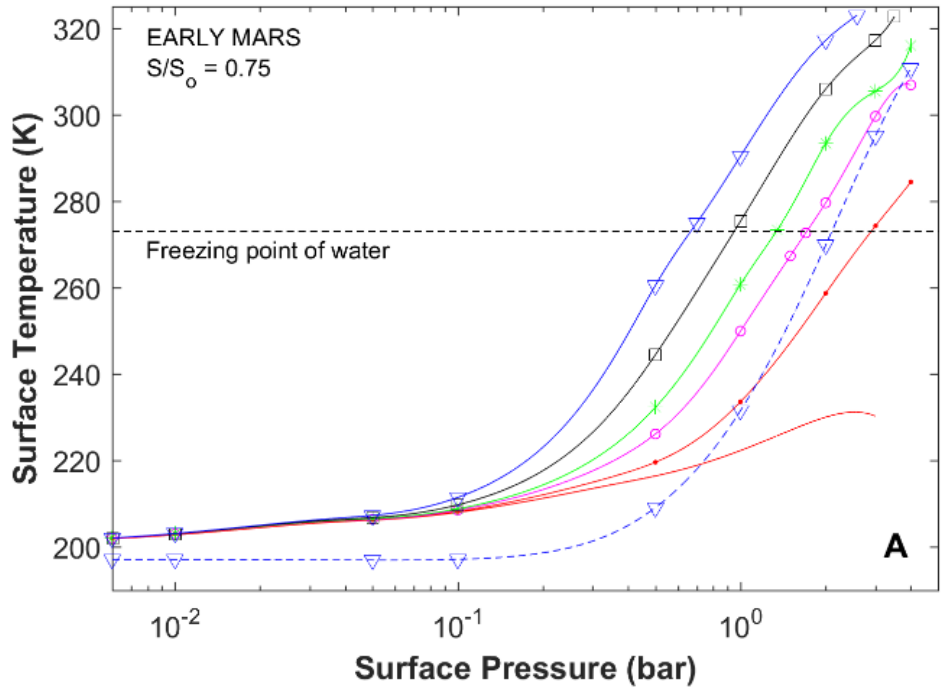

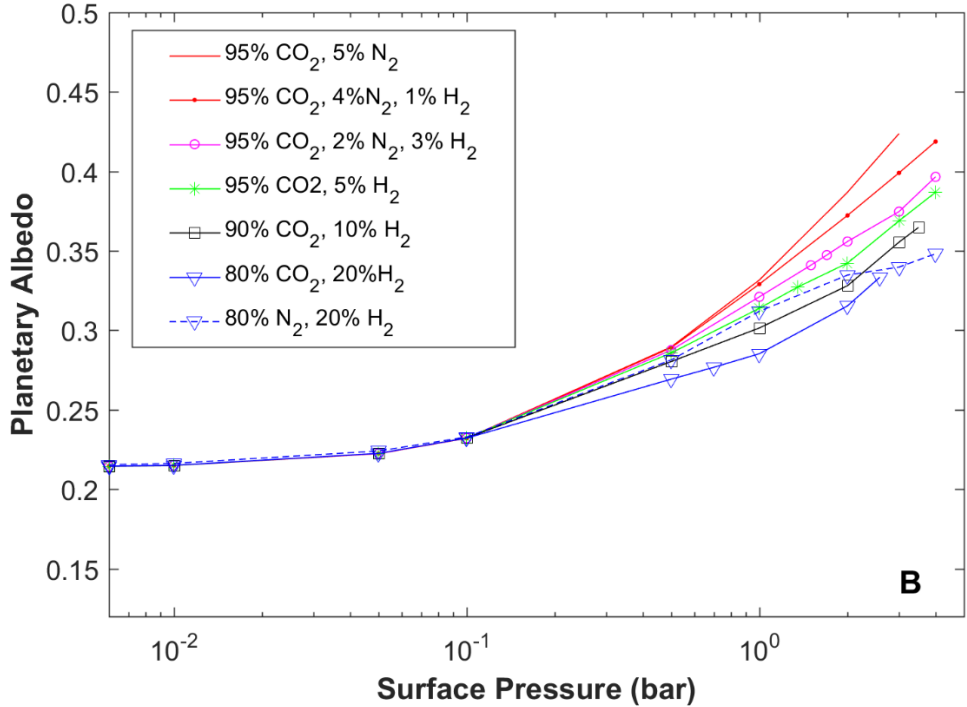

**Figure 1:** Surface temperature (a) and planetary albedo (b) as a function of surface pressure for different fully-saturated atmospheric compositions. The assumed solar luminosity is 0.75 times present, appropriate for 3.8 Ga. The surface albedo is 0.216.

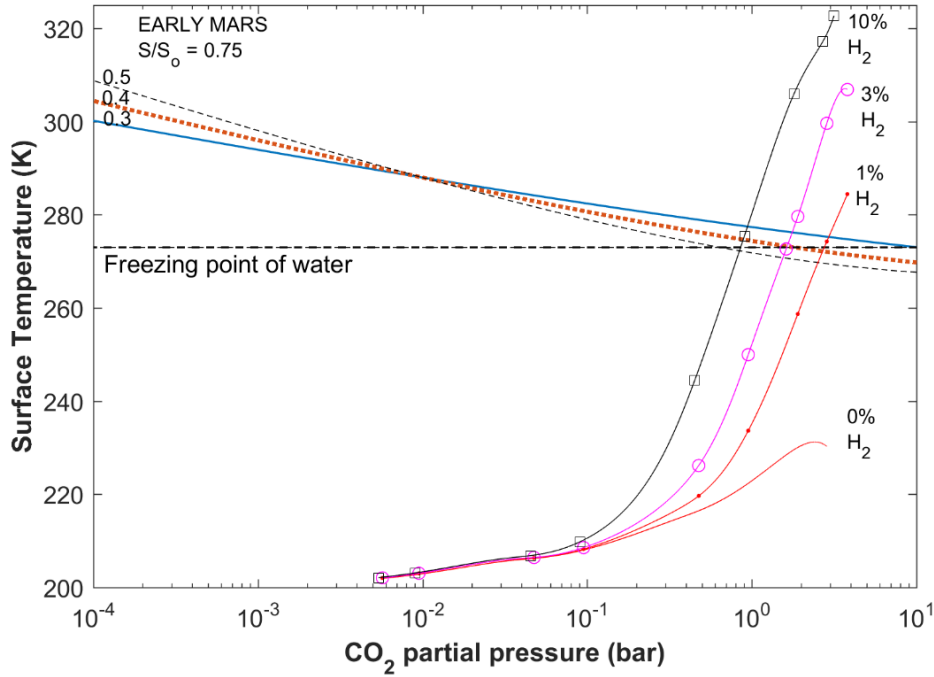

**Figure 2:** Surface temperature as a function of $CO_2$ partial pressure for some of the atmospheric compositions in Figure 1 with weathering rate curves for β values of 0.3 (solid blue), 0.4 (red dots), and 0.5 (black dashed). Limit cycles occur when the temperature and weathering rate curves intersect below the freezing point. However, if the intersection occurs above the freezing point, the climate is stable and persistently warm. The surface remains permanently glaciated if the curves do not intersect. The assumed solar luminosity is 0.75 times present, appropriate for 3.8 Ga. The assumed surface albedo is 0.216.

*3.3 Glaciated early Mars surface albedo study*

The surface albedo assumed in Figures 1 and 2 (0.216) is consistent with a warm, relatively ice-free, early Mars. In contrast, a cold icy early martian surface should have a correspondingly higher surface albedo, requiring higher greenhouse concentrations to warm it. I have revisited the 3%, 10%, and 20% $H_2$ $CO_2$-dominated atmospheres of Figure 1 and repeated the calculation assuming different surface albedo values (Figures 3 – 5). At a modest increase to 0.3, the surface pressure required to achieve above-freezing mean surface temperatures only rises from ~1.7 to 2 in the 3%, ~1 to 1.1 in the 10%, and ~0.7 to 0.75 bar in the 20% $H_2$ cases. However, at still higher surface albedo values, the surface pressure requirement significantly increases. At a surface albedo of 0.4, the required surface pressure for 3% $H_2$ rises to ~2.5 bar (Figure 3a). At a surface albedo of 0.45, the atmosphere collapses for the 3% $H_2$ case as over half the solar energy is reflected to space (Figure 3b). The corresponding pressure for the 10% and 20% $H_2$ cases increases to 1.3 and 0.9 bar, respectively (panels a Figure 4 - 5); for a 0.5 surface albedo the pressure increases to 1.4 bar for the 10% and to 0.95 bar for the 20% $H_2$ cases (panel a Figure 4-5). At a 0.55 surface albedo the atmosphere also collapses for the 10% $H_2$ case (Figure 4b). The greenhouse effect is strongest for the 20% $H_2$ scenario so its atmosphere collapses at a higher surface albedo value (0.65; Figure 5b). The surface pressure in the 20% $H_2$ case increases to ~ 1 bar and ~ 1.1 bar for surface albedo values of 0.55 and 0.6, respectively (Figure 5a). Thus, before collapsing, the pressures required for warm conditions increase by (1.7 – 2.5 bar) 47%, (1 - 1.4 bar) 40%, and (0.7 – 1.1 bar) 57% for the 3%, 10% and 20% $H_2$ cases, respectively. Stable solutions for the 3%. 10%, and 20% $H_2$ scenarios occur for surface albedo values no higher than 0.4, 0.5, and 0.6, respectively. Subsequently, at a surface albedo of 0.3 or higher, the pressures required to establish warm conditions for the 3% $H_2$ scenario exceed available constraints (Kite et al 2014; Hu et al., 2015).

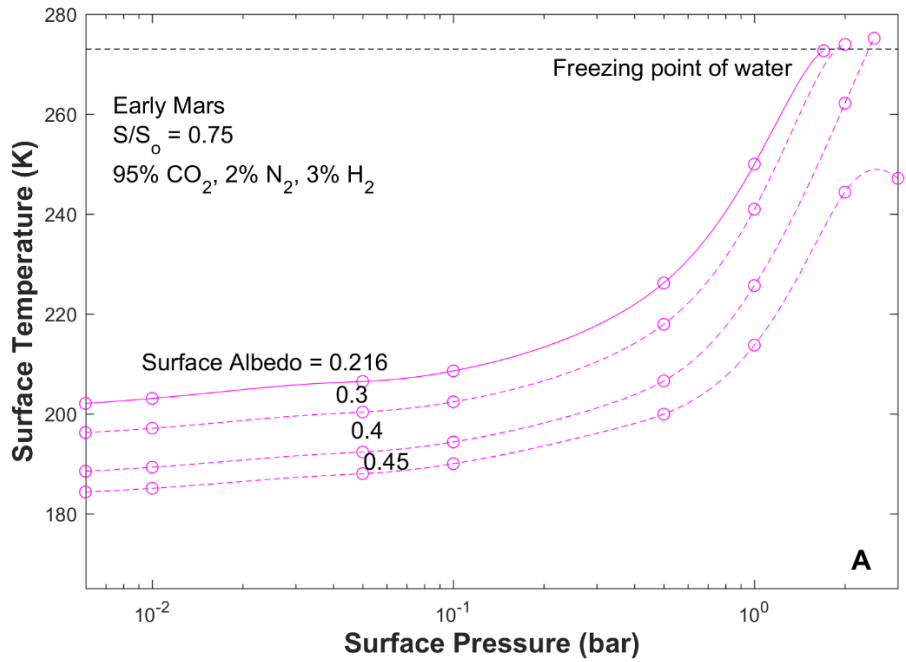

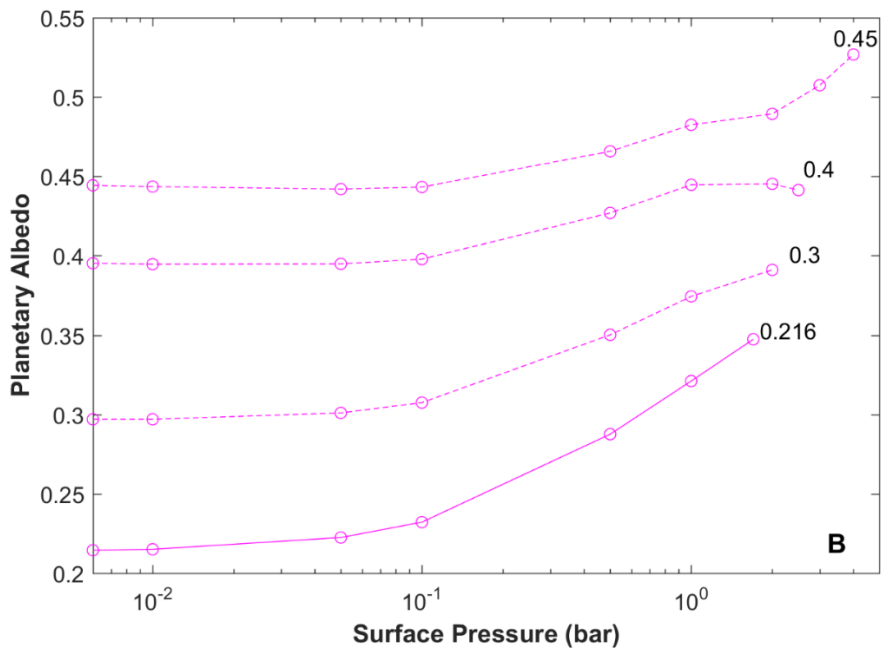

**Figure 3:** Surface temperature as a function of (a) surface pressure and (b) planetary albedo for 95% $CO_2$ and 3% $H_2$ fully-saturated atmosphere from Figure 1 (solid purple line with circles) recalculated at different ice-cover fractions simulated as increases in surface albedo (0.3, 0.45, 0.50, 0.55).

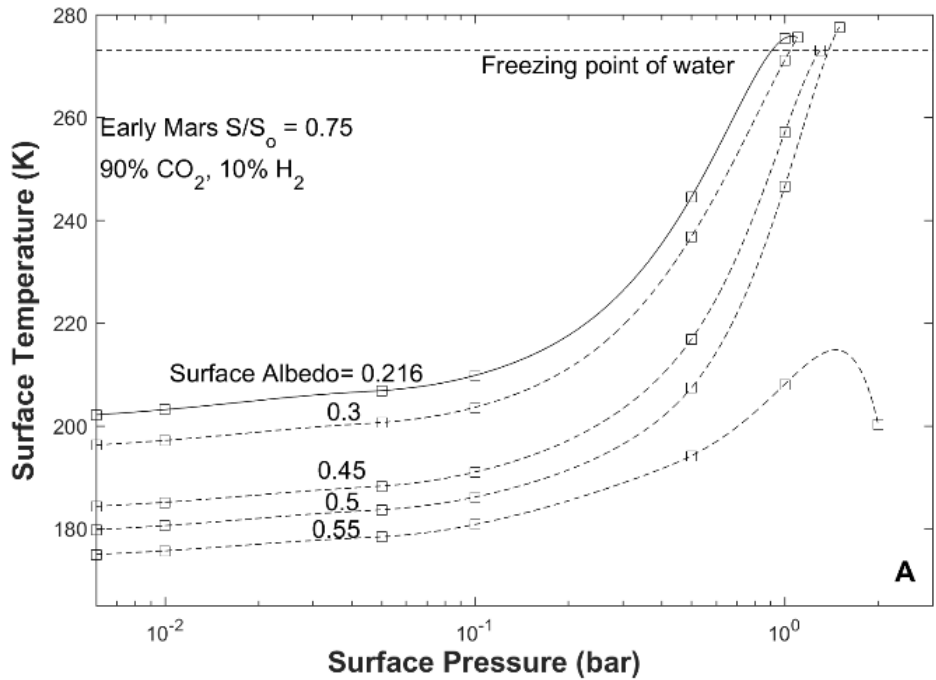

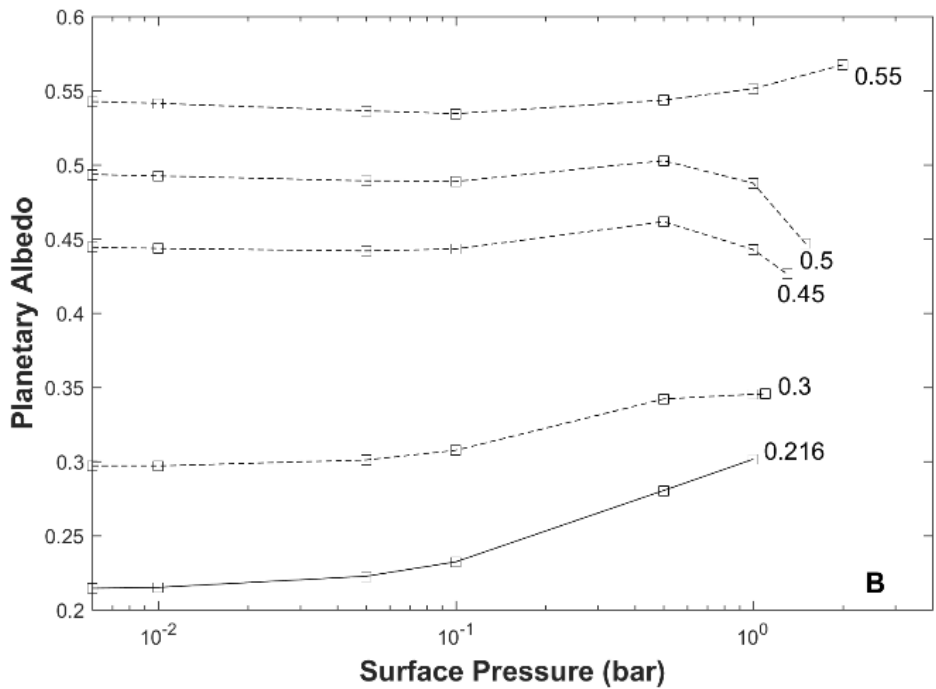

**Figure 4:** Same as in Figure 3 but for the 90% $CO_2$ and 10% $H_2$ fully-saturated atmosphere from Figure 1 (solid black line with squares) recalculated at different ice-cover fractions simulated as increases in surface albedo (0.3, 0.45, 0.50, 0.55).

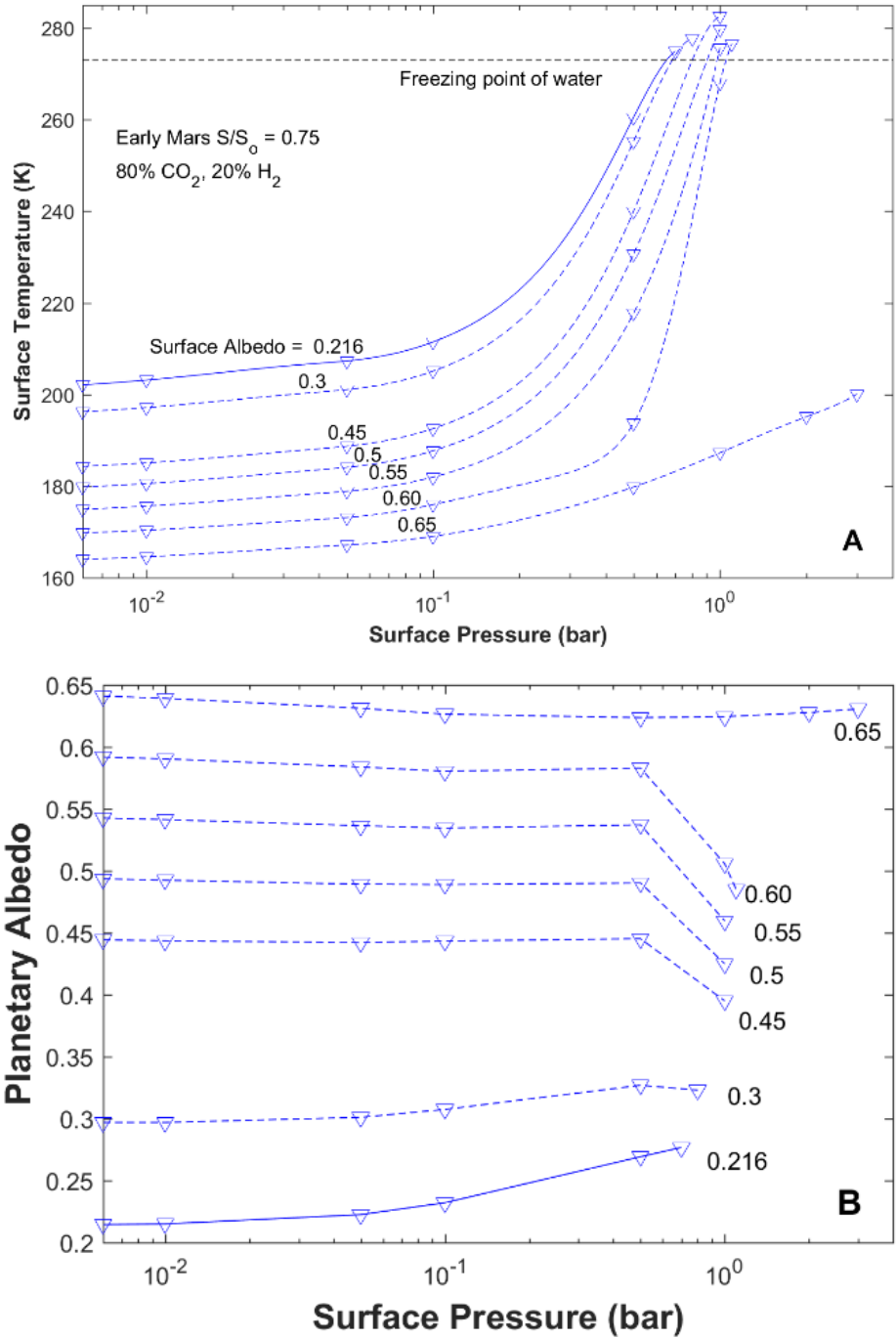

**Figure 5:** Same as in Figure 3 but for the 80% $CO_2$ and 20% $H_2$ fully-saturated atmosphere from Figure 1 (solid blue line with diamonds) recalculated at different ice-cover fractions simulated as increases in surface albedo (0.3, 0.45, 0.50, 0.55, 0.60, 0.65).

*3.4 Cirrus Cloud study*

I utilized the 1 bar 95% $CO_2$ 5% $H_2$ atmosphere of Figure 1 to investigate whether water vapor cirrus clouds can increase its nearly warm mean surface temperature of 261 K to the above-freezing mean surface temperatures (273 K) necessary to produce enough water to form the older valleys and other fluvial features (Figure 6). To provide the additional 12 K and achieve radiative-convective equilibrium, clouds would have to provide ($F_S$ - $F_{IR}$ = ~87.9 – 76.4) ~12.5 W/m² of extra forcing (Figure 7a).

Cirrus cloud decks located between the ~0.33 and 1 bar pressure levels cool the climate further (negative cloud forcing) irrespective of cloud fraction (Figure 7a). Below the 0.33 bar pressure level, clouds warm at all cloud fractions although 25% and 38% cloud cover never provide sufficient warming to generate above-freezing mean surface temperatures. Atmospheric pressure levels below ~0.25 and 0.19 bar, respectively, at 100% and 50% cloud cover, provide enough warming to produce temperatures above 273 K (286 K for 100% and 275 K for 50%). In all cases, the planetary albedo decreases at lower pressure levels, allowing the infrared greenhouse effect to dominate (Figure 7b), a trend that also agrees with observations by Choi and Ho (2006).

Ramirez and Kasting (2017) had found that cirrus cloud fractions exceeding ~75% would have been necessary to warm early Mars, assuming a 0.5 bar $CO_2$-$H_2O$ atmosphere. The denser atmosphere considered here benefitted from the additional heating provided by $H_2$, decreasing the cirrus cloud coverage (50%) needed for warm conditions.

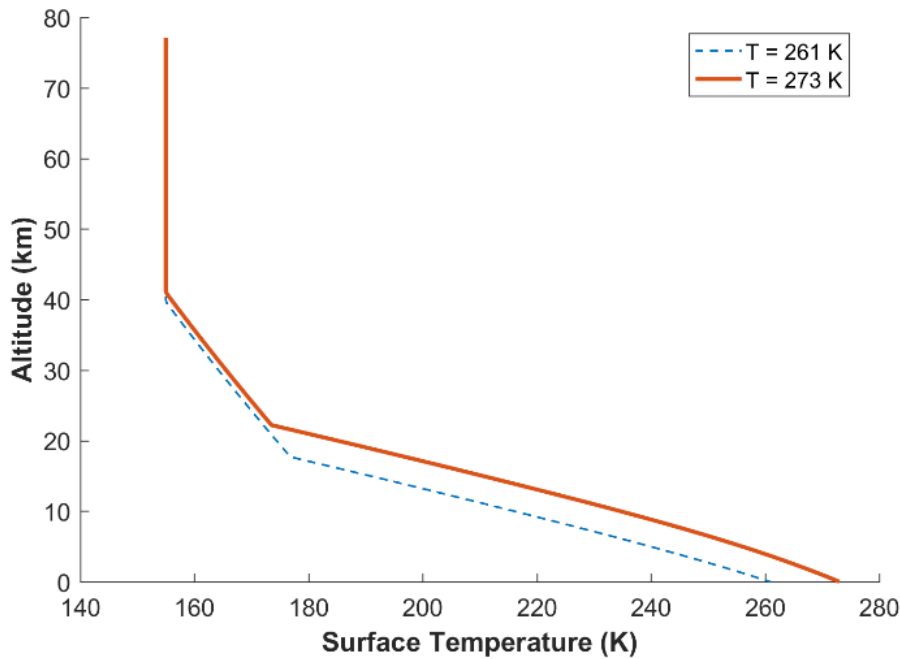

**Figure 6:** Vertical atmospheric profiles for a 1 bar 95% $CO_2$ and 5% $H_2$ atmosphere in radiative-convective equilibrium with a 261 K mean surface temperature (blue dashed line) and that for the target 273 K mean surface temperature profile (solid red line). The assumed solar luminosity is 0.75 times present, appropriate for 3.8 Ga. The surface albedo is 0.216.

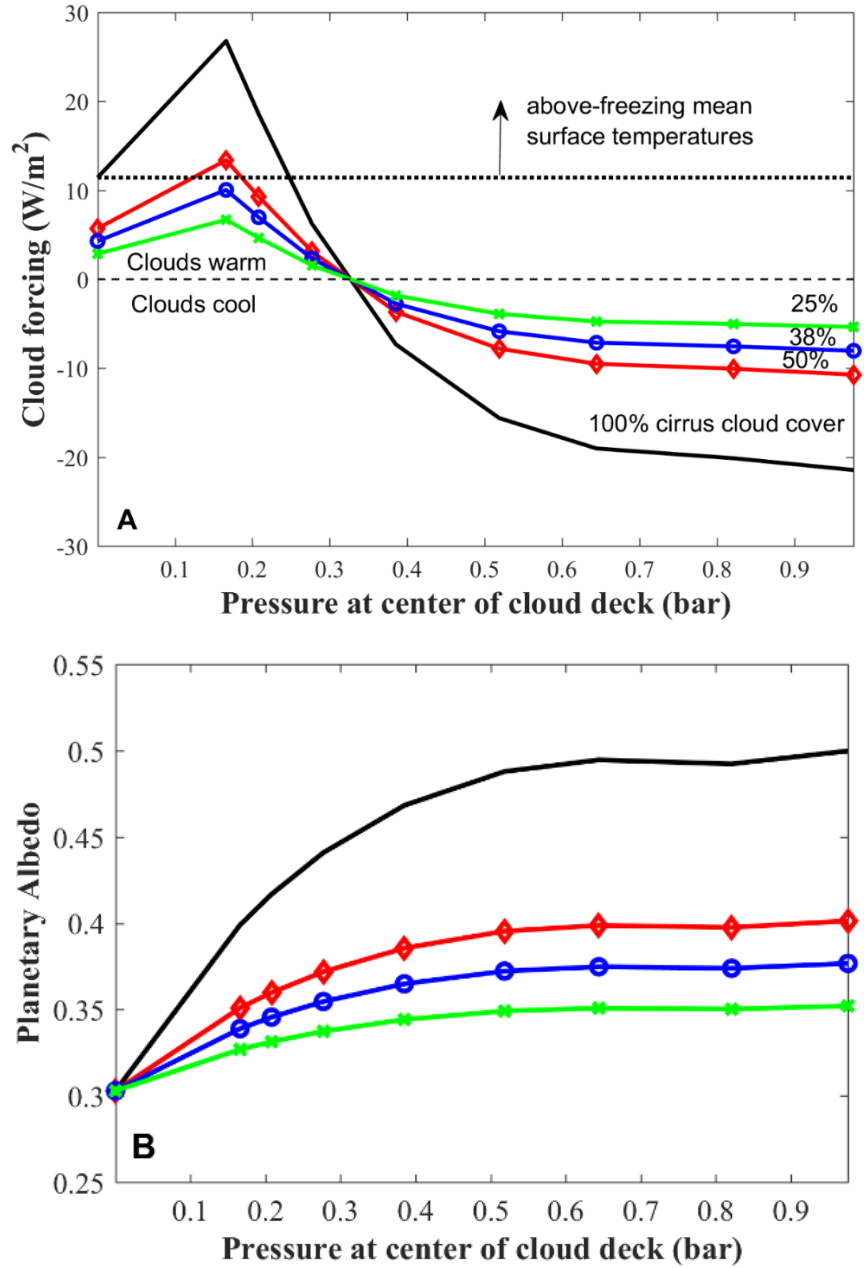

**Figure 7:** Effect of a single 1-km thick cloud layer on (a) cloud forcing and (b) planetary albedo for the 273 K atmosphere in Fig. 6 for 100% (black line), 50% (red line with diamonds), 38% (blue line with circles) and 25% (green line with xs) cloud cover. Cloud forcing that is equal to or greater than that needed to reach 273 K (dotted line) yields warm solutions. The horizontal axis shows the pressure at the center of the assumed cloud deck.

## 3.5 Sensitivity Studies

### 3.5.1 Vertical fluxes as a function of surface albedo

I have compared the incoming solar, outgoing infrared, and net fluxes for the 0.5 bar atmosphere with 3%, 10%, and 20% $H_2$ in Section 3.3 for selected surface albedo values (Figures 8 – 10). As the surface albedo increases (i.e. by increasing surface ice coverage), more solar energy is reflected out to space (Figures 8- 10 panel a), reducing the amount available to warm the troposphere and surface (Figures 8 - 10 panel c). Likewise, both upward and downward thermal emission decreases at higher surface albedo, which leads to a weaker greenhouse effect (Figures 8 – 10 panel b). Note that decreased tropospheric absorption for the cooler, higher surface albedo simulations allows slightly more energy to reach the surface (Figures 8 –10 panel a) although this effect is far outweighed by both the greater net incoming solar energy (Figures 8 – 10 panel a) and enhanced atmospheric emission at thermal infrared wavelengths (Figures 8 – 10 panel b) for the lower surface albedo simulations. The sum of the net incoming solar and net outgoing fluxes as a function of height reveals that the total energy received by the lower atmosphere (pressures > ~0.2 bar) increases as the surface albedo decreases, resulting in higher mean surface temperatures (Figures 8 – 10 panel c).

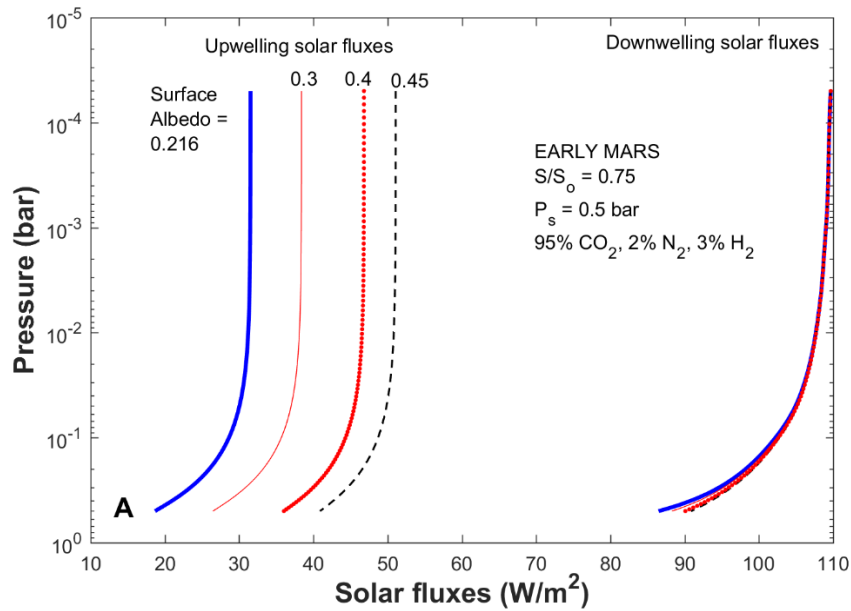

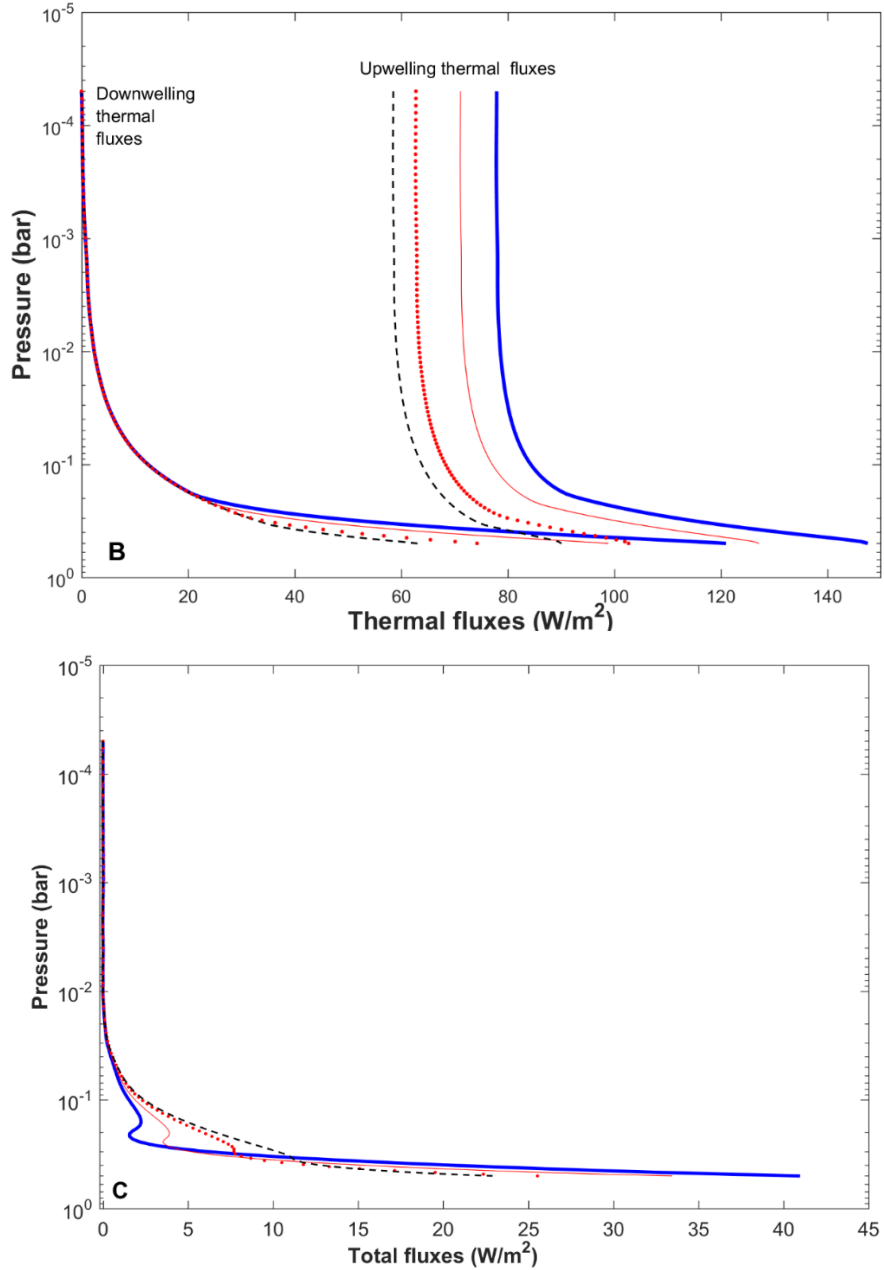

**Figure 8:** Atmospheric pressure versus vertical (a) solar fluxes, (b) thermal fluxes, and (c) total fluxes for the fully-saturated 95% $CO_2$ and 3% percent $H_2$ atmosphere from Figure 1 recalculated at different ice-cover fractions (surface albedo values: 0.216, 0.3, 0.4, 0.45). Total fluxes are obtained by summing the upwelling and downwelling fluxes of each panel and then adding the resultant net sum of thermal and solar fluxes together. The calculated mean surface temperatures (in order of increasing albedo) are ~226, 218, 207, and 200 K, respectively.

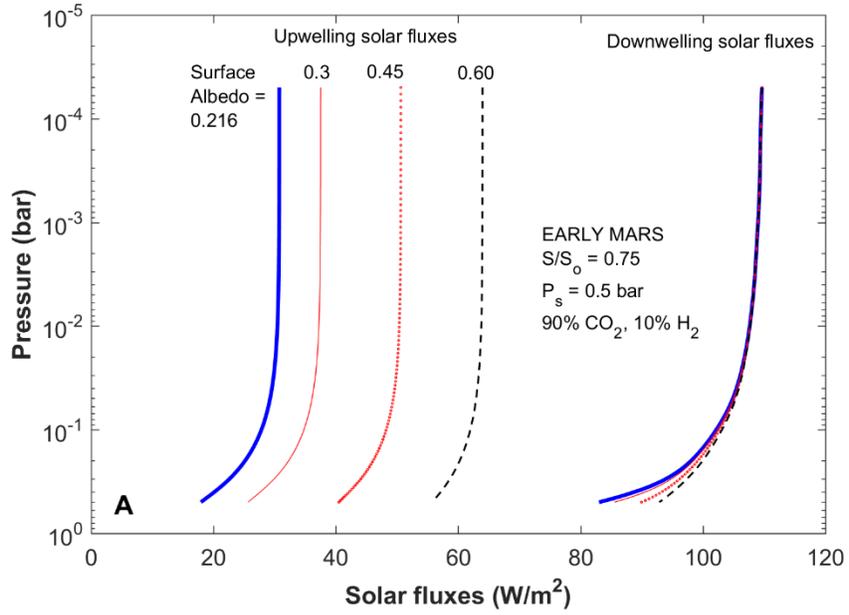

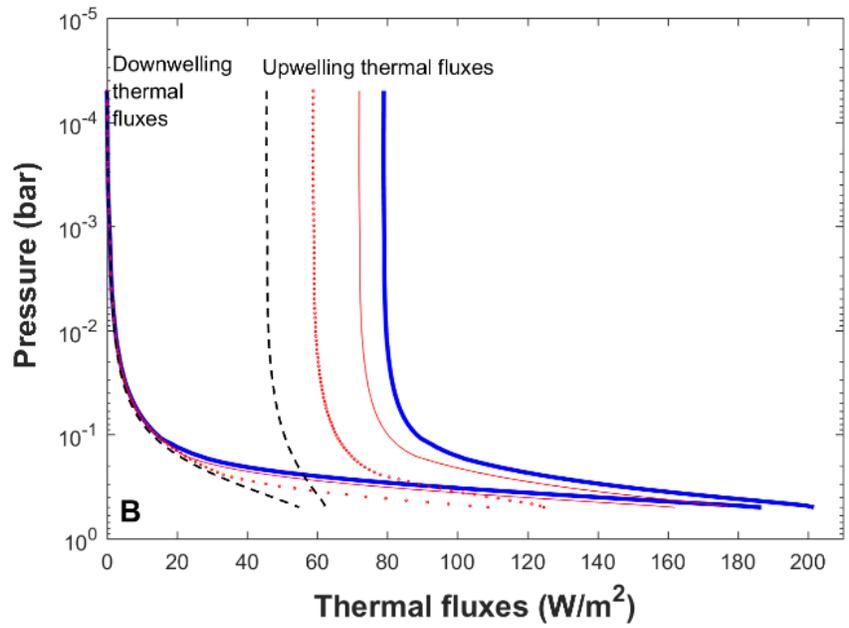

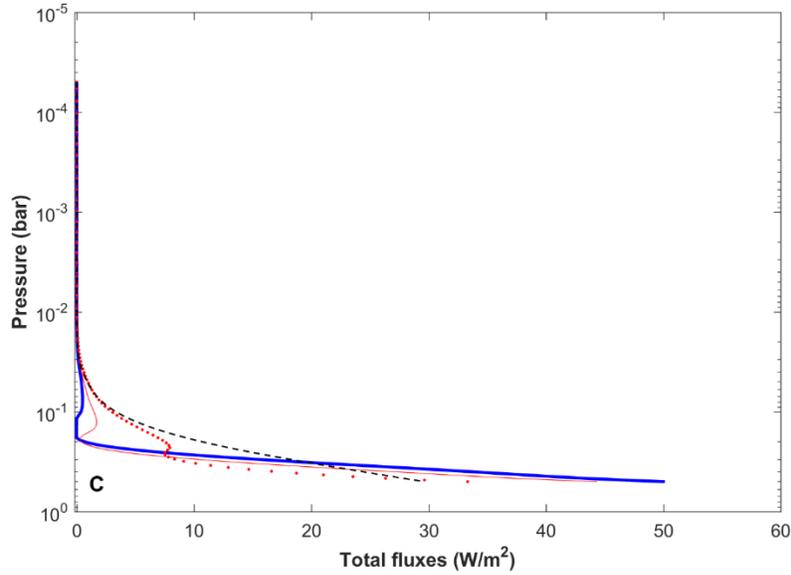

**Figure 9:** Same as figure 8 for the fully-saturated 0.5 bar 90% $CO_2$ and 10% percent $H_2$ atmosphere from Figure 1 recalculated at different ice-cover fractions (surface albedo values: 0.216, 0.3, 0.45, 0.60). The calculated mean surface temperatures (in order of increasing albedo) are ~245, 237, 217, and 182 K, respectively.

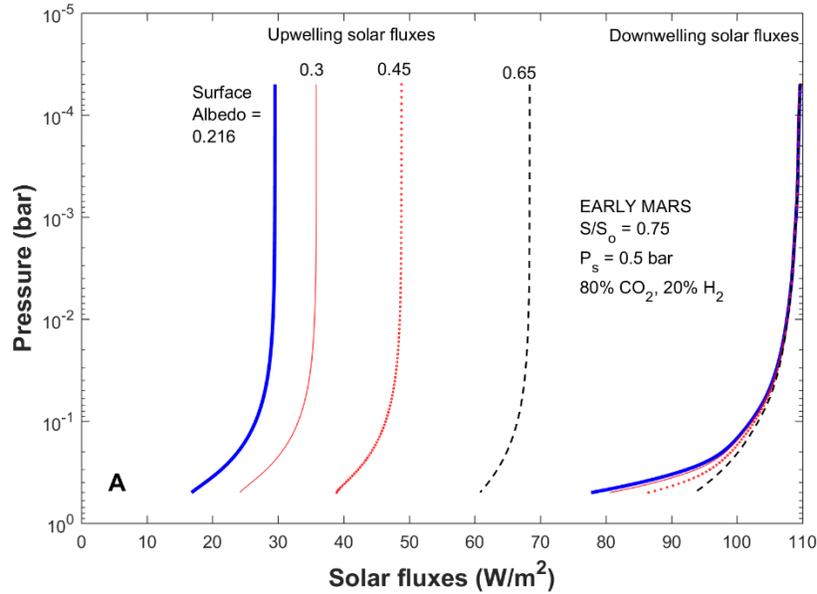

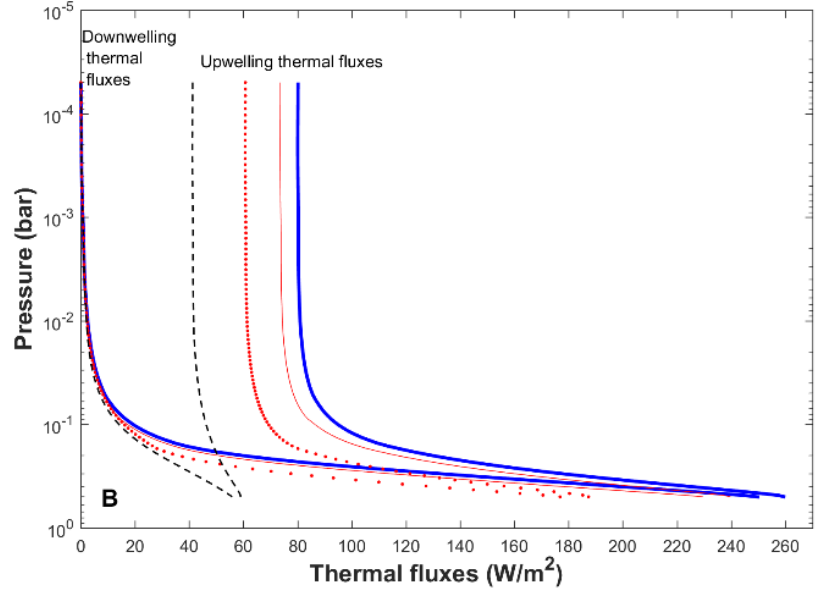

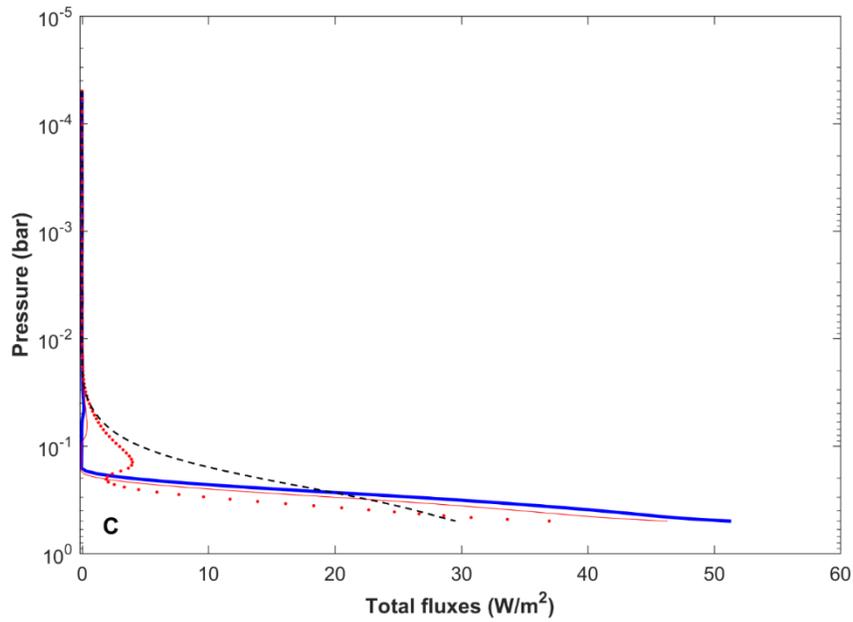

**Figure 10:** Same as figure 8 for the fully-saturated 80% $CO_2$ and 20% percent $H_2$ atmosphere from Figure 1 recalculated at different ice-cover fractions (surface albedo values: 0.216, 0.3, 0.45, 0.65). The calculated mean surface temperatures (in order of increasing albedo) are ~261, 255, 240, and 180 K respectively.

*3.5.2 Emission spectrum for a warm early Mars $CO_2$-$H_2$ atmosphere*

As discussed previously (Ramirez et al. 2014a; Ramirez and Kaltenegger, 2017), foreign-broadening by the background atmosphere (predominantly $CO_2$ in this case) excites $H_2$ roto-translational bands, increasing greenhouse warming in spectral regions where $CO_2$ and $H_2O$ absorb poorly (Figure 11). This effect is much more pronounced than in previous calculations that had assumed $N_2$-$H_2$ CIA as a proxy for $CO_2$-$H_2$ CIA (compare with Figure S6 in Ramirez et al., 2014a).

*3.5.3 Methane sensitivity study*

I have repeated the calculation in Figure S10 of Ramirez et al. (2014a) (Figure 12). A $CH_4$ concentration of 1% is a reasonable upper bound based on a balance of volcanic outgassing and escape rates (e.g. Pavlov 2000; Kharecha et al., 2005; Ramirez et al., 2014). Although Ramirez et al. (2014a) had originally found that $CH_4$ has a weak greenhouse effect, this changes once $CO_2$-$CH_4$ collision-induced absorption is introduced (Figure 12) (Wordsworth et al., 2017 Supporting Info. Figure 3). Moreover, at the resultant higher mean surface temperature, the moist adiabat region expands, decreasing the size of the $CO_2$ cloud deck region. Upper stratospheric heating is still very pronounced. These results suggest that $CO_2$ cloud formation in a warm early martian atmosphere containing appreciable amounts of $CH_4$. would have been significantly suppressed.

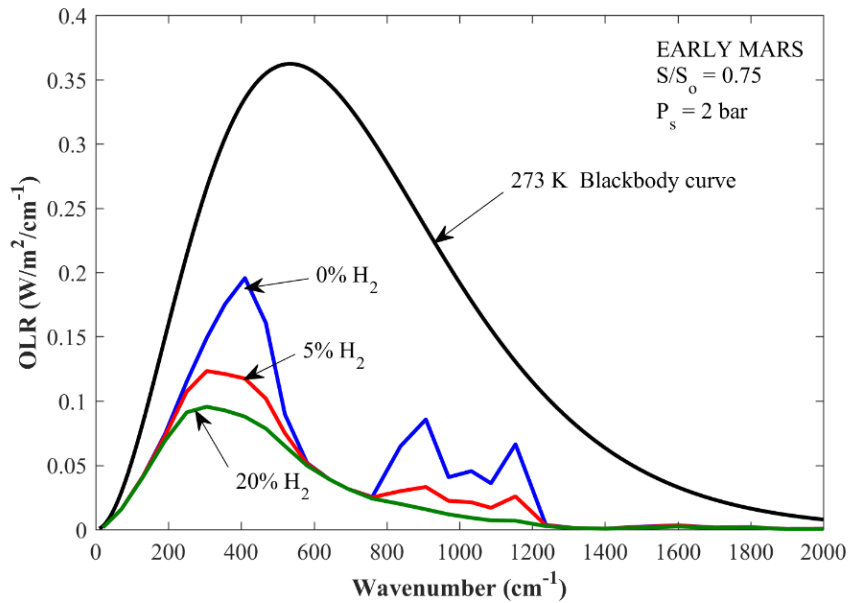

**Figure 11:** Emission spectra for a 2-bar early Mars (S/So = 0.75) atmosphere containing 95% $CO_2$ and 5% $N_2$ (blue), 95% $CO_2$ and 5% $H_2$ (red), or 80% $CO_2$ and 20% $H_2$ (green). The surface temperature is 273 K and the stratospheric temperature is fixed at 167 K. Adding 5% and 20% $H_2$ reduces the outgoing infrared flux from 90.5 W/m$^2$ to ~64 W/m$^2$ and ~50.4 W/m$^2$, respectively. For comparison, adding 5% and 20% $H_2$, $N_2$-$H_2$ CIA only reduces the outgoing infrared flux in Ramirez et al. (2014a) to ~84 W/m$^2$ and 68 W/m$^2$, respectively.

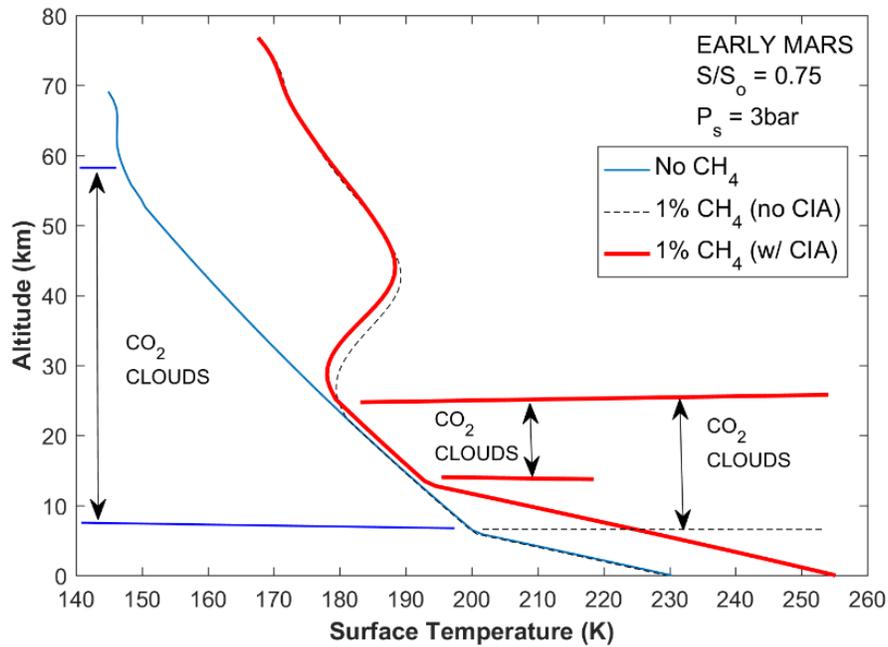

**Figure 12:** Temperature-altitude profiles for fully-saturated 3-bar $CO_2$ early Mars ($S/S_o = 0.75$) atmospheres with 1% $CH_4$ and no collision-induced absorption (black dashed line), 1% $CH_4$ with $CO_2$-$CH_4$ collision-induced absorption (solid red line), and no $CH_4$ (blue line).

# 4. DISCUSSION

*4.1 Difficulties with transient warming considering mineralogy and climate*

In light of the geologic evidence (e.g. Craddock and Howard, 2002), most contemporary arguments against a persistently warm and wet climate have generally transitioned from arguing for permanently cold conditions towards acknowledging that early Mars must have been at least episodically warm. In recent years, a slew of transient warming solutions have been proposed, including seasonal warming (Kite et al. 2013), $CH_4$ and/or $H_2$ bursts resulting from serpentinization processes (e.g. Chassefierre et al. 2013; 2016), glacial melting from icy highlands sheets (e.g., Wordsworth et al. 2013), transient warming from sporadic $SO_2$-rich volcanic eruptions (Halevy and Head, 2014), impact-induced cirrus clouds (Urata and Toon, 2013), impact-induced runaway greenhouses (Segura et al., 2002; 2008, 2012), and limit cycles for planets near the outer edge of the habitable zone (Batalha et al. 2016).

Arguments in favor of such episodic warming scenarios have either focused on the presence of minerals (e.g. olivine, jarosite) that would not have lasted long in a persistently warm climate or perceived minimal chemical alteration inferred to be suggestive of cold conditions. (Clark and Hoefen, 2000; Madden et al. 2004; Tosca and Knoll, 2009; Wordsworth, 2016). However, timing issues plague all of these interpretations. Either the detected minerals were not formed during the period of valley network formation (in some case billions of years later, e.g. Soderblom, 1992), or the terrains themselves were not formed during older valley formation. The latter is the case of the jarosite found in Meridiani Planum (e.g., Madden et al. 2004; Tosca and Knoll, 2009), a region that is likely even more ancient than the older valleys themselves (Squyres et al. 2004). More importantly, the discovery of opaline silica occurs alongside other minerals in older terrains, suggesting considerable alteration must have occurred (Carter et al. 2012; 2013).

My climate solutions do not readily support transient warming solutions for the late Noachian-early Hesperian time frame either. According to Bishop et al. (2017), elevated temperatures would enable formation of the observed abundance of clay minerals under short (< ~tens of thousands of years; Janice Bishop, personal communication) transient episodes through faster reaction rates, requiring surface temperatures between ~298 – 323 K. This may be especially problematic if such transient episodes were few in number. In the most optimistic 20% $H_2$ case, surface pressures exceeding 2.6 (1.3) bar satisfy the requirement to reach a mean surface temperature of 323 K (298 K) (Figure 1). In the 5% $H_2$ case, surface pressures must exceed 5 (2.5) bar at 323 K (298 K). At still lower $H_2$ concentrations, surface pressures may need to exceed 10 bar. With the exception of the 298 K 20% $H_2$ case, all of these pressures well exceed available constraints (Kite et al. 2014; Hu et al., 2015). Note that these surface pressures are still underestimates and the actual required surface pressures would be even higher on a cold icy Mars (see below). In contrast, comparatively lower reaction rates, and substantially lower pressures (see Results), would suffice to achieve warm mean surface temperatures well below 300 K and produce the abundance of observed surface materials in a persistently warm climate.

*4.2 Transient warming and the ice problem*

Moreover, these transient scenarios inherently assume an initially cold early Mars. However, if Mars had begun cold the implication is a largely glaciated planetary surface, with ice accumulation being centered in polar regions and the southern highlands (e.g. Fastook and Head, 2012; Wordsworth et al. 2013). Indeed, 3-D simulations suggest that the global ice coverage on a cold early Mars may have exceeded ~25 - 30% (Wordsworth et al. 2013), which would have significantly increased the surface albedo above the nominal value (~ 0.2) and made warm solutions even more difficult to achieve.

I estimate the surface albedo of a cold and glaciated early Mars by making conservative assumptions about surface ice. If I assume a typical value for snow/ice mixtures (0.65) then warm solutions are obtainable for ice coverage no higher than ~40%, 70%, and 85% for the 3%, 10%, and 20% $H_2$ cases, requiring surface albedo values no higher than ~0.4, 0.5, and 0.6, respectively (panel b Figures 3 – 5). For fresher ice/snow (assume albedo = 1), warm solutions are impossible for ice coverage above ~25%, 35%, and 49 % for the 3%, 10%, and 20% $H_2$ cases, respectively. Although episodically warm solutions are possible if the ice was dirty enough (so long as global coverage did not exceed ~40%), my results show that surface pressures required to achieve warm conditions on a glaciated early Mars can be ~ 40 – 60% higher (see Section 3.3). This ice problem compounds that explained in Section 4.1 if temperatures exceeding ~300 K are required to form surface clay materials over a limited number of episodes (Bishop et al. 2017).

*4.3 Lack of evidence for an early glaciated Mars*

The geologic evidence does not support a heavily glaciated early Mars either (e.g. Craddock and Howard, 2002). Glacial features, such as cirques, kames, and eskers, are noticeably absent in ancient terrains (e.g. Wordsworth, 2016). Furthermore, the geomorphology strongly suggests that a widespread process, most likely precipitation, was the major erosive agent on early Mars (e.g. Craddock and Howard, 2002; Davis et al. 2016), countering the notion that localized sources of glacial melt from icy highland sheets (e.g. Wordsworth et al. 2013; 2015) could have formed these fluvial features, including those in Arabia Terra (Davis et al. 2016). Indeed, a non-glaciated and warm early Mars is consistent with the lack of observed glacial features in these early terrains (e.g., Grotzinger et al.,, 2015).

*4.4 The effect of cirrus cloud warming in a $CO_2$-$H_2$ atmosphere*

Could cirrus clouds significantly change the interpretations here? These results do not indicate this. Previous simulations (Forget et al. 2013) suggest that the *total* cloud fraction on early Mars was ~50%, which is similar to that on the Earth (Ramirez and Kasting, 2017). This is because regions of rising moist air must balance those of subsiding dry air (Bartlett and Hunt, 1972). However, cirrus cloud fractions exceeding ~75% in a $CO_2$-$H_2O$ atmosphere (Ramirez and Kasting,

2017) or 50% in one buttressed by secondary gases (this study) are likely to be too high unless clouds had operated very differently on early Mars. Moreover, the additional presence of low water clouds would have cooled the climate, indicating that these estimated minimum cirrus cloud fractions may be underestimates. For instance, clouds have an overall cooling effect on the Earth although cirrus clouds having a warming effect (e.g. Ramanathan et al., 1989). Nevertheless, I show that thin (1-km) cirrus clouds composed of 100-micron particles located at the most favorable pressure level (~0.17 bar, Figure 7a), with typical terrestrial cirrus cloud coverage fractions (25 – 38%) (e.g. Wylie et al. 1994) in a fully-saturated atmosphere, are still insufficient to provide enough warming, even in the absence of low clouds that would cool the climate.

*4.5 The effects of obliquity on polar ice production and the onset of atmospheric collapse*

Obliquity effects are important because they can impact the production of surface ice and influence the onset and location of atmospheric collapse. Such effects cannot be accurately determined with single-column climate models and so it is relevant to discuss recent 3-D results and their potential implications for this work. Wordsworth et al. (2013) showed that low obliquity conditions (25 degrees) favor the formation of northern polar $H_2O$ ice caps in addition to those formed in the southern hemisphere. At higher obliquity (45 degrees), however, increased insolation in the northern hemisphere would melt those $H_2O$ ice caps. The formation of these caps raises the surface albedo, which greatly increases the difficulty of achieving warm conditions as this study shows.

Similarly, $CO_2$ polar ice is favored to form at low obliquities, which if produced in sufficient amounts, triggers atmospheric collapse (e.g. Forget et al., 2013; Soto et al. 2015). Thus, $CO_2$ glaciation is even more harmful to the maintenance of warm climates than is $H_2O$ glaciation because not only does the surface albedo increase, but the $CO_2$ mole fraction decreases as well, which reduces both the $CO_2$ greenhouse effect itself and the accompanying water vapor feedback. Forget et al. (2013) found that atmospheric collapse for early Mars occurs at pressures below 1 bar at low obliquity and for pressures exceeding 3 bar at high obliquity. Analogously, when $CO_2$ condensation outstrips the greenhouse effect in a 1-D model, a maximum temperature is achieved and further increases in $pCO_2$ often lead to temperature decreases (Figs. 1 -5) (see also Kasting, 1991; Ramirez et al., 2014a). That said, the current evidence suggests that such high atmospheric pressures (3 bar) during the late Noachian/Hesperian were probably never reached, at least not persistently (Kite et al., 2014; Hu et al., 2015). Thus, the applicability of such high pressures to the early Mars problem is unknown.

Moreover, Soto et al. (2015) found that decreased meridional heat transport resulting from the interaction with Olympus Mons and the Tharsis Montes region could trigger atmospheric collapse at still lower pressures than those found in previous work (e.g. Forget et al, 2013). Interestingly, they also find that at high enough $CO_2$ pressures, the meridional transport is intense enough to outpace the effects of $CO_2$ condensation, eliminating the formation of $CO_2$ polar ice caps. The Soto et al. analysis should be repeated for early Mars conditions to confirm this result.

Nevertheless, these studies (Forget et al., 2013; Wordsworth et al., 2013; Soto et al. 2015) have only assessed obliquity in the context of a cold planet. Indeed, the possibility of atmospheric collapse on a planet with a chaotic obliquity cycle is yet another reason why transiently warming an icy early Mars is so challenging. In contrast, an initially warm early Mars powered by the $CO_2$-$H_2$ greenhouse mechanism discussed here would have enhanced meridional transport and be far less susceptible to atmospheric collapse or the formation of polar caps (be they $CO_2$ or $H_2O$), even at low obliquity. The effects of $CO_2$ condensation prevent the 2 bar $CO_2$ ($H_2$-free) case in Figure 1 from achieving mean surface temperatures exceeding 230 K. However, the addition of just 1% $H_2$ enhances the greenhouse effect sufficiently to allow mean surface temperatures to exceed the freezing point of water (Figure 1). Thus, these results suggest that even modest amounts of hydrogen could greatly reduce, and possibly eliminate, the likelihood of atmospheric collapse. This prediction should be checked at both high and low obliquity using models that include both meridional heat transport and a latitudinal dependence. Future work should also assess the effects of eccentricity because it has a big effect on the seasonal cycle and could impact, not just the onset of collapse (e.g. Soto et al. 2015), but also the degree of surface melting (e.g. Kite et al. 2013).

*4.6 A warm and semi-arid early climate*

It is possible that early Mars was characterized by a warm semi-arid climate that produced enough precipitation (e.g. rainfall) to form the ancient valleys (e.g. Craddock and Howard, 2002). Such an interpretation is consistent with the observed lack of glaciation in ancient terrains (Craddock and Howard, 2002; Wordsworth 2016), the water amounts necessary to form the older valleys (Hoke et al. 2011), and the recent estimates for a smaller initial water inventory (< 200 m global equivalent) (Villanueva et al., 2015). Such a small initial water inventory also explains why Mars may have never experienced global glaciation nor the snowball episodes that appear to have occurred on the early Earth (Hoffman et al., 1998).

A $CO_2$-$H_2$ greenhouse with $H_2$ concentrations exceeding ~3%, satisfies available paleopressure and climate stability constraints (Kite et al., 2014; Hu et al., 2015; Batalha et al. 2016). A warm early Mars supplied by volcanism is likely the most efficient manner of achieving such concentrations. One criticism is that high $CO_2$ concentrations may be difficult to achieve with a reduced mantle (e.g. Wordsworth et al. 2017). However, as had been argued previously (Wetzel et al., 2013; Ramirez et al.,2014a; Batalha et al., 2015), high $CO_2$ concentrations would have been indirectly produced as $H_2O$ vapor reacted with outgassed reduced gases, like CO and $CH_4$, in addition to what may have survived primordially.

Although scant detections of serpentine have been observed from orbit (Ehlmann et al., 2010), there is no evidence that serpentinization (e.g. Chassefierre et al. 2013; 2016), a process by which Fe-rich waters produce $H_2$ via oxidation of basaltic crust, has ever occurred on the global extent necessary to produce such concentrations. Moreover, transient warming via serpentinization is challenged by both the ice problem and perhaps by, elevated surface temperatures and pressures necessary to form observed surface materials (Bishop et al. 2017).

Nevertheless, it is possible that transient warming did occur later in Mars' history when it was significantly drier and colder. It is the author's opinion that transient warming mechanisms in frigid climates are more appropriate for the late Hesperian or Amazonian. These epochs are characterized by smaller fluvial features and sedimentary deposits, with resultant geomorphologies inconsistent with formation by precipitation (e.g. Gulick and Baker, 1989; 1990). Fluvial features formed in these later periods may not have required a warm climate (e.g. Kite et al. 2013; Bristow et al. 2017).

*4.7 Low Hesperian pCO$_2$ levels?*

A recent study adds an extra complication to Mars paleoclimate studies. Bristow et al. (2017) estimates that pCO$_2$ levels during the formation of the Gale Crater deposits (~3.5 Gya) were in the 10s mb range, much too low for climate models with traditional greenhouse gases to obtain warm conditions. If true, the implications of that analysis are difficult to explain irrespective of the atmospheric history of early Mars. If a thick (> 0.5 bar CO$_2$) atmosphere was required to warm early Mars at ~3.8 – 3.6 Gya, both transiently and persistently warm models would have to address how nearly all of that CO$_2$ could have been lost within as little as 100 million years later. The Gale Crater deposits are also difficult to explain with a persistently cold early Mars scenario. According to Figure 1, mean surface temperatures for early Mars at the 10mb level would have been < 210 K, possibly even colder than today because of the faint young Sun. Such temperatures and pressures are much too low for liquid water to exist in a stable state anywhere on the planetary surface, except transiently and in relatively small quantities (e.g. Fairen et al., 2009). It would therefore seem unlikely that the abundant fluvial evidence seen at Gale Crater, which abounds with m-scale sedimentary deposits and evidence for repeated fluvial flows that had likely fed a long-standing lake (e.g. Grotzinger et al., 2015; Hurowitz et al., 2017), could have been formed under such a cold and dry martian climate. Relatively high sedimentation rates were also likely needed to form alluvial fan deposits during this time frame, suggesting at least quasi-persistently wet surface conditions (Kite et al. 2017). Thus, given the conflicting data and interpretations, more work is needed to verify the Bristow et al. (2017) results and understand their significance for early Mars.

## 5. CONCLUSION

I update the calculations in Ramirez et al. (2014a) and compute the first physically consistent persistently and transiently warm and wet early Mars solutions for $CO_2$-$H_2$ atmospheres. Assuming fully-saturated $CO_2$-dominated compositions, I find that warm conditions under a non-glaciated and warm and wet early climate can be achieved with $H_2$ concentrations as low as 1%. Hydrogen concentrations above just ~3%, with surface pressures between ~ 0.7 bar (0.55 bar $pCO_2$) and 20% $H_2$ or 1.7 bar and ~3% $H_2$, satisfy both available paleopressure and climate stability constraints (Kite et al., 2014; Hu et al. 2015, Batalha et al. 2016). Volcanic outgassing and the indirect products of water vapor photolysis would have maintained such concentrations, assuming an initially reduced martian mantle.

I show that modeled surface pressures should be significantly higher than calculated in Wordsworth et al. (2017) because the surface albedo of a glaciated early Mars would be higher than one that is not. The addition of surface ice in such cold Mars scenarios increases the surface pressures required for warm conditions by ~ 10 – 60%, depending on surface ice coverage. No warm solution is possible for ice cover fractions exceeding 40%, 70%, and 85% for snow/ice and 25%, 35%, and 49% for fresher snow/ice for 3%, 10%, and 20% $H_2$, respectively. Such transient warming scenarios may also require high surface temperatures (298 – 323 K) to produce the observed quantities of surface materials, which may be difficult if the number of such episodes is small. Surface pressures to achieve mean surface temperatures of 323 K exceed available paleopressure constraints for all $H_2$ concentrations considered (1 – 20%) except that a 298 K mean surface temperature can be achieved at a surface pressure of 1.3 bar with 20% $H_2$.

I confirm recent work (Wordsworth et al. 2017) showing that $CH_4$ has a net greenhouse effect once $CO_2$-$CH_4$ collisions are incorporated. The addition of $H_2$ produces warm solutions with just 50% cirrus cloud cover although this still exceeds the expected cirrus cover for warm planets with both moist rising and dry subsiding air (e.g. Ramirez and Kasting, 2017).

I predict that atmospheric collapse from $CO_2$ condensation should be difficult to achieve in warm early Mars scenarios with $H_2$ mixing ratios that exceed ~1 or 2%. This should be tested with models that have meridional heat transport and include a latitudinal dependence. Future work should also assess different cloud particle shapes (i.e. fractals) for both water vapor and $CO_2$ clouds and evaluate how the cloud greenhouse effect changes from the spherical particle assumption.

## ACKNOWLEDGEMENTS

The author would like to thank Bob Craddock, Janice Bishop, and James F. Kasting for the helpful comments and discussions. The author also thanks the two anonymous referees for their feedback. The author acknowledges support by the Simons Foundation (SCOL # 290357, Kaltenegger), Cornell Center for Astrophysics and Planetary Science, and the Carl Sagan Institute.